\documentclass[fleqn,usenatbib]{mnras}

\usepackage{newtxtext,newtxmath}

\usepackage[T1]{fontenc}

\DeclareRobustCommand{\VAN}[3]{#2}
\let\VANthebibliography\thebibliography
\def\thebibliography{\DeclareRobustCommand{\VAN}[3]{##3}\VANthebibliography}

\usepackage{graphicx}	
\usepackage{amsmath}	

\title[Constraint on the meVSL using TD in SNeIa]{Constraint on the minimally extended varying speed of light using time dilations in type Ia supernovae}

\author[S. Lee]{
Seokcheon Lee $^{1}$\thanks{E-mail: skylee@skku.edu}
\\
$^{1}$Department of Physics, Institute of Basic Science, Sungkyunkwan University, Suwon 16419, Korea }

\date{Accepted XXX. Received YYY; in original form ZZZ}

\pubyear{2023}

\begin{document}
\label{firstpage}
\pagerange{\pageref{firstpage}--\pageref{lastpage}}
\maketitle

\begin{abstract}
The Friedmann-Lema\^{i}tre-Robertson-Walker model establishes the correlation between redshifts and distances. It has a metric expansion of space. As a result, the wavelength of photons propagating through the expanding space is stretched, creating the cosmological redshift, $z$. It also relates the frequency of light detected by a local observer to that emitted from a distant source. In standard cosmology (\textit{i.e.}, a constant speed light model),  this relation is given by a factor $1/(1+z)$.  However, this ratio is modified in the minimally extended varying speed of light model (meVSL, $c = c_0 a^{b/4}$) as $1/(1+z)^{1-b/4}$.  This time dilation effect is detected as the observed rate of the time variation in the intensity of emitted radiation. The spectra of type Ia supernovae (SNe Ia) provide a reliable way to measure the apparent aging rate of distant objects. We use data on 13 high-redshift ($0.28 \leq z \leq 0.62$) SNe Ia to obtain $b = 0.198 \pm 0.415$ at the $1$-$\sigma$ confidence interval. The current data is too sparse to give meaningful constrain on the meVSL and cannot distinguish the meVSL model from the standard model. 
\end{abstract}




\section{Introduction}

The spectrum of light coming from a distant source shows features in it, such as absorption lines, emission lines, or other variations in light intensity, and one can use them to determine the (cosmological) redshift \citet{Harrison:2000}.  Redshift characterizes the relative difference between observed and emitted wavelengths of an object.  Friedmann's solutions to Einstein's equations of general relativity (GR) by adopting the Robertson-Walker (RW) metric show that the wavelength of photons propagating through the expanding space is stretched, creating the redshift. Thus, the redshift, $z$, is the fundamental observable in the Friedmann-Lema\^{i}tre-Robertson-Walker (FLRW) model.

The stretching of the wavelength of photons in the expanding universe can be related to the shrinking of the frequency by using the dispersion relation of photons.  Consequently, this effect should be detected as the time dilation (TD) in the intensity of emitted radiation \citet{2019LRR....22....1I}. 

The type Ia supernovae (SNe Ia) light evolutions have been used as cosmic clocks to measure TD in distant objects due to their large luminosities and variability on short timescales. Both light curves (LCs) of distant SNe Ia  \citet{1996ApJ...466L..21L,2001ApJ...558..359G} and their spectral evolutions \cite{2008ApJ...682..724B,1997AJ....114..722R,2005ApJ...626L..11F} provide methods to measure the apparent aging rate of faraway objects.  Especially the latter removes the degeneracy between the intrinsic LC width and TD effect to make it more reliable.  

In standard cosmology (SC \textit{i.e.}, $c =$ constant),  the ratio of the frequency of light detected by a local observer to that emitted from a distant source is given by a factor $1/(1+z)$ \cite{Weinberg:1972}.  However, this ratio is modified in the minimally extended varying speed of light model (meVSL, $c = c_0 a^{b/4}$) as $1/(1+z)^{1-b/4}$ \cite{2021JCAP...08..054L,2023FoPh...53...40L,2023MNRAS.522.3248L}. Thus, one can use the TD effect detected in SNe Ia to distinguish these models.  

We organize this paper as follows. In section \ref{sec:redshift}, we review the TD formula of the meVSL model. In section \ref{sec:analysis},  we analyze it with 13 high-redshift ($0.28 \leq z \leq 0.62$) SNe Ia \cite{2008ApJ...682..724B} data to constrain the meVSL model.  We give our conclusion in section \ref{sec:Conc}.

\section{Review of cosmological redshift}
\label{sec:redshift}

In the expanding universe, one can obtain it by defining the proper distance $d_{\textrm{p}}$ as the multiplication of the comoving distance $\chi$ by the scale factor $d_{\textrm{p}} \equiv a(t) \chi$.  Comoving distance is the distance between two points measured along a path defined at the present cosmological time.  It remains constant in time for objects moving with the Hubble flow. From this definition, one obtains the relation between proper distances at $t_1$ and $t_2$ as
\begin{align}
\chi = \frac{d_{\textrm{p}}(t_1)}{a(t_1)} =  \frac{d_{\textrm{p}}(t_2)}{a(t_2)} \quad \Rightarrow \quad d_{\textrm{p}}(t_1) = \frac{a(t_1)}{a(t_2)} d_{\textrm{p}}(t_2) \label{r} \,.
\end{align}
Now one can replace $d_{\textrm{p}}(t_1)$ as the measured wavelength at present ($a(t_0) = 1$) and $d_{\textrm{p}}(t_2)$ as the emitted one in the past ($a(t_\textrm{emit})$) to find their relationship with redshift, $z$ as
\begin{align}
\lambda_{\textrm{obs}} = \frac{1}{a(t_\textrm{emit})} \lambda_{\textrm{emit}} \equiv (1+z)  \lambda_{\textrm{emit}} \quad \textrm{where} \quad z \equiv \frac{\lambda_{\textrm{obs}} - \lambda_{\textrm{emit}}}{\lambda_{\textrm{emit}}} \label{lambdaobs} \,.
\end{align}

One can relate redshift $z$ to the frequency in the RW model given by
\begin{align}
ds^2 &= g_{\mu\nu} dx^{\mu} dx^{\nu} \nonumber \\
&= -\left( dx^0\right)^2 + a(t)^2 \left( \frac{dr^2}{1 - k r^2} + r^2 \left( d \theta^2 + \sin^2 \theta d \phi^2 \right) \right) \nonumber \\
&\equiv -\left( dx^0\right)^2  + a(t)^2 \left( d \chi^2 + r^2 d \Omega^2 \right) \label{ds2} \,,
\end{align}
where $x^{0} = ct$, $r$ is the coordinate distance, and $d \Omega$ is the solid angle.  To derive the redshift,  one uses the geodesic equation for a light wave $ds^2 = 0$.  Let us assume that an observer at a position $r=0$ and time $t = t_0$ measures the first crest of a light wave emitted at a position $r = R$ and time $t = t_{\textrm{emit}}$. At the time $t_0 + \delta t_0$, the observer sees the next crest of the same light emitted at a time $t_{\textrm{emit}} + \delta t_{\textrm{emit}}$. Then, one obtains the following equation from the RW metric 
\begin{align}
\chi &= \int_{0}^{R} \frac{dr}{\sqrt{1-kr^2}} = \int_{t_{\textrm{emit}}}^{t_0} \frac{dx^0}{a(t)} = \int_{t_{\textrm{emit}}+ \delta t_{\textrm{emit}}}^{t_0+ \delta t_0} \frac{dx^0}{a(t)} \nonumber \\
&\Rightarrow \int_{t_0}^{t_0+\delta t_0} \frac{dx^0}{a(t)} = \int_{t_{\textrm{emit}}}^{t_{\textrm{emit}}+ \delta t_{\textrm{emit}}} \frac{dx^0}{a(t)} \label{chi} \,.
\end{align}
For tiny variations in time (throughout one cycle of a light wave),  scale factors are essentially constants (\textit{i.e.}, $a(t_{\textrm{emit}}+ \delta t_{\textrm{emit}}) \approx a(t_{\textrm{emit}})$ and $a(t_0+\delta t_0) \approx a(t_0)$). It yields 
\begin{align}
& \frac{1}{a(t_{0})} \Bigl[ x_0(t_{0}+ \delta t_{0}) -  x_0(t_{0}) \Bigr] \nonumber \\
&= \frac{1}{a(t_{\textrm{emit}})} \Bigl[ x_0(t_{\textrm{emit}}+ \delta t_{\textrm{emit}}) -  x_0(t_{\textrm{emit}}) \Bigr] \label{zx0} \,.
\end{align} 
In SC,  $x_0(t_{0}+ \delta t_{0}) -  x_0(t_{0}) = c \delta t_0$ and $x_0(t_{\textrm{emit}}+ \delta t_{\textrm{emit}}) -  x_0(t_{\textrm{emit}}) = c \delta t_{\textrm{emit}}$ to make Eq.~\eqref{zx0} as
\begin{align}
 \frac{c \delta t_0}{a(t_{0})}  = \frac{ c \delta t_{\textrm{emit}}}{a(t_{\textrm{emit}})} \quad \Rightarrow \quad \lambda(t_{0})= \frac{1}{a(t_{\textrm{emit}})} \lambda (t_{\textrm{emit}}) =  (1+z)  \lambda_{\textrm{emit}} \label{lambdaobsSCM} \,,
\end{align}  
where we define the frequency as $1/\delta t_i = \nu (t_i) = \nu(t_0) a(t_i)^{-1}$ and use the relation $c = \lambda (t_i) \nu (t_i) = $ constant \cite{Weinberg:1972}.  Eq.~\eqref{lambdaobsSCM} is identical to Eq.~\eqref{lambdaobs}. In the meVSL model (\textit{i.e.}, $c(t_i) = c_0 a(t_i)^{b/4}$) and $x_0(t_{0}+ \delta t_{0}) -  x_0(t_{0}) \approx c_0 a(t_0)^{b/4} \delta t_0$, thus $x_0(t_{\textrm{emit}}+ \delta t_{\textrm{emit}}) -  x_0(t_{\textrm{emit}}) \approx c_0 a(t_{\textrm{emit}})^{b/4}  \delta t_{\textrm{emit}}$. From this fact, Eq.~\eqref{zx0} becomes
\begin{align}
& \frac{c_0 \delta t_0}{a(t_{0})^{1-b/4}}  = \frac{ c_0 \delta t_{\textrm{emit}}}{a(t_{\textrm{emit}})^{1-b/4}} \nonumber \\ &\Rightarrow \lambda(t_{0})= \frac{1}{a(t_{\textrm{emit}})} \lambda (t_{\textrm{emit}}) =  (1+z)  \lambda_{\textrm{emit}} \label{lambdameVSL} \,,
\end{align}  
where we use the frequency as $1/\delta t_i = \nu (t_i) = \nu(t_0) a(t_i)^{-1+b/4}$ and use the relation $c(t_i) = \lambda (t_i) \nu (t_i)$ \cite{2000ApJ...532L..87B,2014PhLB..728...15B,2021JCAP...08..054L,2023FoPh...53...40L}.  Again one can recover the redshift defined in the expanding universe even in the meVSL model shown in equation ~\eqref{lambdameVSL}. Also, some other VSL models have provided the same equation for the redshift of the radiation wavelength, $\lambda$, and the scale factor, $a$, as in SC \cite{2020MNRAS.498.4481G,2023MNRAS.519..633C}. 

\section{Analysis}
\label{sec:analysis}

In an expanding universe, the galaxies are all moving away from each other, and clocks run at different rates to produce the TD in a distant galaxy. In the RW metric, the space is homogeneous and isotropic, and the expansion is so regular that the world lines of the galaxies form a nonintersecting and diverging three-bundle of geodesics. One can define spacelike hypersurfaces that are orthogonal to all of them, and one can label each hypersurface by a constant value of the global cosmic time coordinate $t$ \citet{Islam:2001,Roos:2015}. 

However, the rate of local clocks depends on the hypothesis of the constancy of the speed of light as shown in the previous section \ref{sec:redshift}.  Thus,  it is required to compare models with current TD data to clarify the underlying model.  However,   there are few available observational data due to the difficulty of the direct measurement of TD and its independence of cosmological parameters. Fortunately, LCs of distant SNe Ia provide data for TD \citet{1996ApJ...466L..21L,2001ApJ...558..359G}, and more reliable data is available from spectral evolutions of a high redshift SNe Ia \citet{1997AJ....114..722R,2005ApJ...626L..11F,2008ApJ...682..724B}. 

\subsection{Brief review of SNe as clocks}
\label{subsec:SNeclocks}

The LC of an SN tracks its brightness over time, showing how its luminosity changes as it evolves. It begins with a rise in brightness called the "pre-maximum" phase, reaching peak luminosity, followed by a decline. The shape of the LC reveals details about the SN, especially for SNe Ia used as standard candles in cosmology. By studying the LC, astronomers determine peak brightness, time to reach maximum brightness, and decline rate, aiding the classification and understanding of energetics, composition, and explosion mechanisms. Comparing LCs at different distances allows investigation of cosmic expansion and TD, leading to discoveries like accelerating universe expansion and dark energy. 

Wilson's method, proposed in 1939, compares the LCs of nearby and distant SNe \cite{1939ApJ....90..634W}. Those of distant SNe exhibit TD, appearing stretched compared to nearby ones due to the travel time of light through space. By analyzing the time dilation effect in LCs, one can determine the time it took for light to reach the observer from SNe at different distances. This information helps study the expansion rate of the Universe and test cosmological models. Practically, this involves gathering data on supernova brightness and evolution at various redshifts, fitting mathematical models to their LCs, and comparing observed TD with model predictions. 

However, in SNe Ia, there is an inherent correlation between the breadths of their LCs and their peak luminosities. Brighter SNe Ia tend to have broader LCs \cite{1993ApJ...413L.105P}. It can lead to confusion when distinguishing between the effects of time dilation and variations in luminosity, especially if the latter is perceived as a stretching of the LC.

The temporal evolution of SN Ia spectra offers a more reliable approach to measuring the passage of time for individual SN. One can use the wavelengths of permanent spectral features to determine the timing of a supernova's peak brightness \cite{1939ApJ....89..156M}. The evolution of spectral characteristics in most SNe Ia follows a consistent pattern, allowing for an objective assessment of a supernova's age based on its spectrum. Comparing two or more spectra of the same distant supernova to those of nearby SNe measures the rate at which the distant supernova ages. This method of measuring the aging rate provides a direct means to verify the presence of time dilation expected in an expanding Universe \cite{1997AJ....114..722R}.

\begin{table*}
\centering
\begin{tabular}{|cccc|}
\hline
SN & $z$ & $1/(1+z)$ & aging rate (error)  \\
\hline 
1996bj & 0.574 & 0.635 & 0.527 (0.369) \\
1997ex & 0.361 & 0.735 & 0.745 (0.076) \\
2001go & 0.552 & 0.644 & 0.652 (0.062) \\
2002iz & 0.427 & 0.701 & 0.655 (0.089) \\
b027   & 0.315 & 0.760 & 0.823 (0.092) \\
2003js & 0.363 & 0.734 & 0.718 (0.082) \\
04D2an & 0.621 & 0.617 & 0.567 (0.341) \\
\hline
\end{tabular}
\begin{tabular}{|cccc|}
\hline
SN & $z$ & $1/(1+z)$ & aging rate (error)  \\
\hline 
2006mk & 0.475 & 0.678 & 0.753 (0.060) \\
2006sc & 0.357 & 0.737 & 0.619 (0.121) \\
2006tk & 0.312 & 0.762 & 0.835 (0.181) \\
2007tg & 0.502 & 0.666 & 0.687 (0.102) \\
2007tt & 0.374 & 0.728 & 0.718 (0.108) \\
2007un & 0.283 & 0.779 & 0.759 (0.135) \\
& & &  \\
\hline
\end{tabular}
\caption{These are aging rate measurements from SNe Ia observation \citet{2008ApJ...682..724B}. }
\label{tab:table-1}
\end{table*}

\subsection{The minimum $\chi^2$ analysis}
\label{subsec:chi2}

We show aging rate measurements of $13$ high-redshift SNe Ia data provided in the reference \citet{2008ApJ...682..724B}. They are given in Tab.~\ref{tab:table-1}.  They are based on a sample of $35$ spectra of $13$ SNe Ia in the redshift range $0.28 \leq z \leq 0.62$ by comparing their differences between the observer-frame ages and the rest-frame ones. 

\begin{table}
\begin{center}
\begin{tabular}{ |ccccc|} 
 \hline
Model & $\chi^2$/dof & $\bar{b}$ & $1$-$\sigma$ & GoF (\%)  \\ 
\hline
$(1+z)^{-1}$ & $3.6/13$ & $0$ & $0$ & $99.5$ \\ 
$(1+z)^{-1+b/4}$ & $3.4/12$ & $0.198$ & $0.415$ & $99.2$  \\  
\hline
\end{tabular}
\end{center}
\caption{A simple $\chi^2$ analysis for both the SC and the meVSL model using TD data \citet{2008ApJ...682..724B}.  The best-fit value for the $b$ exponent is $b = 0.198$ with $0.415$ for the $1$-$\sigma$ confidence interval.  }
\label{tab:chi2}
\end{table}

We perform a least-square fit to the data for predictions of both the SC and the meVSL model
\begin{align}
\chi^2 = \sum_{i=1}^{13} \frac{\left( (1+z_i)_{\textrm{obs}}^{-1} - (1+z_i)^{-1+b/4} \right)^2}{\sigma_i^2} \label{chi2} \,,
\end{align}
where we use Eq.~\eqref{lambdameVSL}, $ (1+z_i)_{\textrm{obs}}^{-1}$ denote 13 data points, and $\sigma_i$ represent their errors in Table~\ref{tab:table-1}. The SC aging rate $1/(1+z)$ yields a fit to the data ($\chi^2 = 3.6$ for $13$ degrees of freedom, d.o.f) with a $99.5$ \% goodness-of-fit (GoF).  The same $\chi^2$ analysis for the meVSL model provides the best-fit value of parameter $b = 0.198$ with the $1$-$\sigma$ error interval being $0.415$.  The minimum $\chi^2$ value is $3.4$ with $12$ d.o.f.  Its GoF is $99.3$\%. These results are summarized in Table.~\ref{tab:chi2}.    

\begin{figure*}
\centering
\vspace{1cm}
\includegraphics[width=1\linewidth]{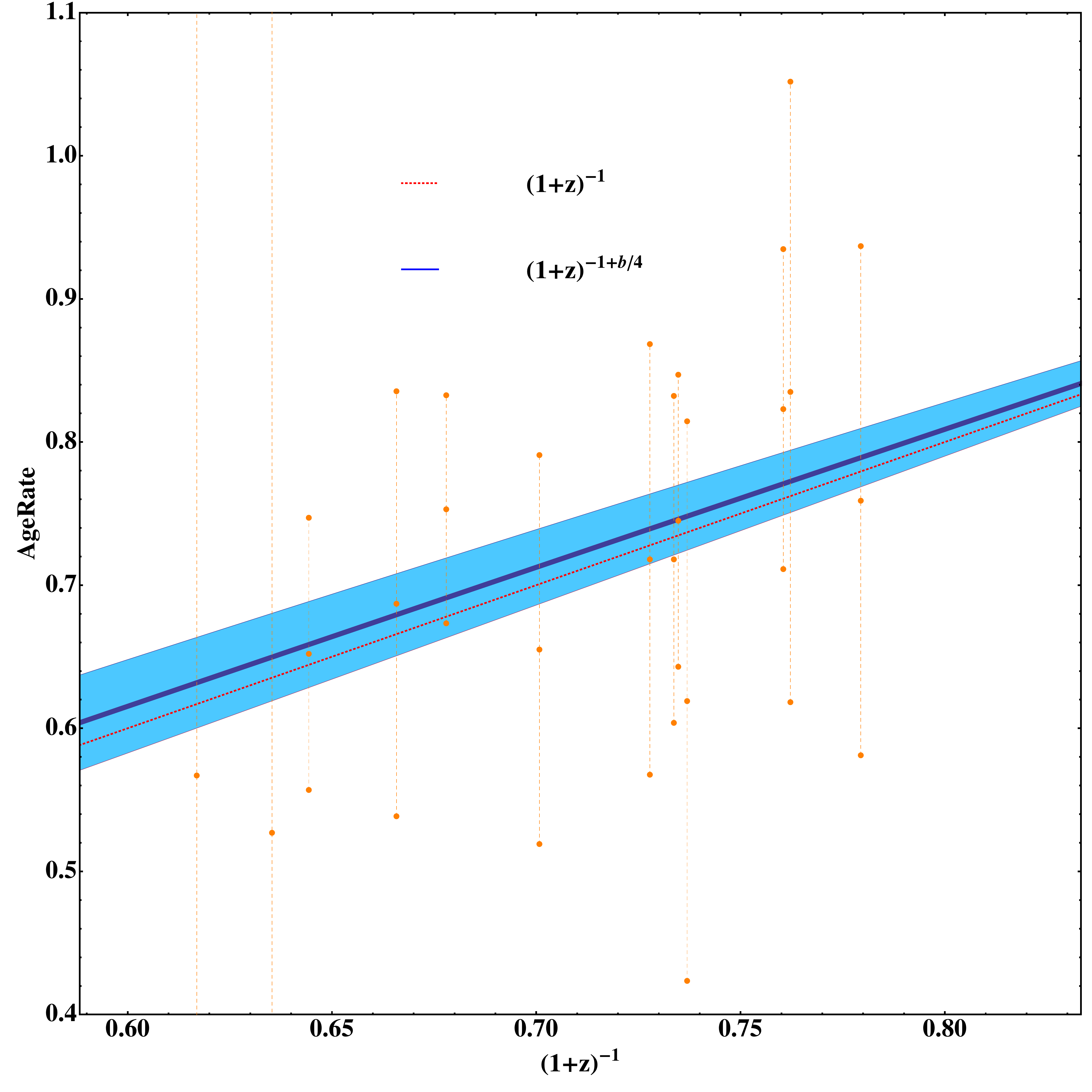} 
\vspace{-0.5cm}
\caption{Time dilation for two models with data points. The dashed line depicts the SC prediction of TD. The thick solid line represents the best-fit value predicted TD of the meVSL model.  The shaded region means the $1$-$\sigma$ confidence interval region. } \label{fig-1}
\vspace{1cm}
\end{figure*}

We also show this in Fig.~\ref{fig-1}.  The vertical lines depict $13$ data points with their measurement errors. The dashed line represents the SC prediction of the aging rate $1/(1+z)$. The solid line with shade regions depicts the meVSL model within the $1$-$\sigma$ confidence interval.  This result shows that the current data cannot distinguish the meVSL from the SC. 

Both the SC and the meVSL have almost the same GoF values. It means that both models provide a similar level of agreement between the observed data and predictions of models. Thus, both models are almost equally preferred to current SNe data. Furthermore, the result from the analysis of the meVSL model cannot be distinguished from that of SC due to its large error.  However, there are about $1500$ SNe LCs available in the redshift range $0 \leq z \leq 2.3$ from Pantheon22 data \cite{2022ApJ...938..113S,2022ApJ...938..110B}. The 1-$\sigma$ error of $b$ from the Pantheon22 data can be $0.01$ if $\sigma_{i}$ is about $0.1$ and $b \sim 0.1$ by using Fisher matrix analysis (This is the preliminary result of our ongoing research).  Thus, if TD data for Pantheon22 data is available, then we might have a chance to distinguish the meVSL model from the SC. 

\section{Conclusion}
\label{sec:Conc}

We have driven a formula for the time dilation of a distant object in the context of the minimally extended varying speed of the light model $T(z) = T_0 (1+z)^{1-b/4}$. We obtain the best-fitting values for the b exponent as $b=0.198\pm0.415$ at the $1$-$\sigma$ confidence level from $13$ high-redshift SNe Ia time dilation data. This result is worse than one obtained using the cosmological chronometers \cite{2023MNRAS.522.3248L}. The best-fitting values for the $b$ are $b=-0.105\pm0.178$ and $0.584\pm0.184$ adopting ($H_0,\Omega_{m0}$) values from Planck18 \cite{2020A&A...641A...6P} and Pantheon22 \cite{2022ApJ...938..110B}, respectively.  The result also shows that the current data are consistent with both the standard model expectation and that of the meVSL model, and we cannot distinguish between the meVSL model and the standard one from the time dilation data using SNe.  There is also time dilation data from gamma-ray bursts \citet{2022JCAP...02..010S}. However, the current unbinned GRB data is too sparse to be directly used for investigating models with time dilation. Recently, there is a confirmation of the cosmic time dilation in 190 quasars in the SDSS Stripe 82 region \cite{2023arXiv230604053L} through long-term multi-waveband monitoring spanning over two decades using SDSS, PanSTARRS-1, the Dark Energy Survey, and dedicated follow-up monitoring with Blanco 4m/DECam \cite{2022MNRAS.514..164S}. Using Bayesian analysis, they successfully detect the expected redshift-dependent time dilation in quasar variability as $n \equiv (1-b/4) = 1.28^{+0.28}_{-0.29}$. This result contradicts previous claims of its absence \cite{2010MNRAS.405.1940H}.  Previous failure to detect time dilation in quasar variability may be attributed to limitations such as small sample sizes and insufficient data sampling and characterization. However, a significantly larger and diverse sample of quasars across various redshifts makes it possible to detect time dilation. Thus, we might have a chance to prefer the specific model in future observations even though the current SNe Ia data cannot provide the statistically meaningful result to distinguish models.


\section*{Acknowledgments}
SL is supported by Basic Science Research Programs through the National Research Foundation of Korea (NRF) funded by the Ministry of Science, ICT, and Future Planning (Grant No. NRF-2017R1A2B4011168 and No. NRF-2019R1A6A1A10073079). 

\section*{Data Availability}

The authors confirm that the data supporting the findings of this study are available within the references and their supplementary materials. No new data were created or analyzed in this study.



\bibliographystyle{mnras}
\bibliography{example} 








\bsp	
\label{lastpage}
\end{document}